\documentclass[conference]{IEEEtran}
\IEEEoverridecommandlockouts
\usepackage{cite}
\usepackage{amsmath,amssymb,amsfonts}
\usepackage{algorithmic}
\usepackage{graphicx}
\usepackage{textcomp}
\usepackage{xcolor}
\usepackage{comment}
\usepackage{array}
\usepackage{float}

\def\BibTeX{{\rm B\kern-.05em{\sc i\kern-.025em b}\kern-.08em
    T\kern-.1667em\lower.7ex\hbox{E}\kern-.125emX}}
\begin{document}

\title{MAE-SAM2: Mask Autoencoder-Enhanced SAM2 for Clinical Retinal Vascular Leakage Segmentation \\
}

\author{
Xin Xing$^{\dagger}$$^{*}$\thanks{* Corresponding author.},
Irmak Karaca$^{\ddagger}$, 
Amir Akhavanrezayat$^{\mathsection}$,
Samira Badrloo$^{\dagger}$, 
Quan Dong Nguyen$^{\mathsection}$,  
Mahadevan Subramaniam$^{\dagger}$\\
\\
$^{\dagger}$Department of Computer Science, University of Nebraska Omaha, Omaha, NE, USA \\
$^{\ddagger}$Department of Ophthalmology, Columbia University Irving Medical Center, New York, NY, USA \\
$^{\mathsection}$Department of Ophthalmology, Stanford University, Stanford, CA, USA \\
\texttt{\{xxing, sbadrloo, msubramaniam\}@nebraska.edu}, \texttt{irmakkaracamd@gmail.com},\\
\texttt{\{akhavana, ndquan\}@stanford.edu}
}

\maketitle

\begin{abstract}
We propose MAE-SAM2, a novel foundation model for retinal vascular leakage segmentation on fluorescein angiography images. Due to the small size and dense distribution of the leakage areas, along with the limited availability of labeled clinical data, this presents a significant challenge for segmentation tasks. Our approach integrates a Self-Supervised learning (SSL) strategy, Masked Autoencoder (MAE), with SAM2. In our implementation, we explore different loss functions and conclude a task-specific combined loss. Extensive experiments and ablation studies demonstrate that MAE-SAM2 outperforms several state-of-the-art models, achieving the highest Dice score and Intersection-over-Union (IoU). Compared to the original SAM2, our model achieves a $5\%$ performance improvement, highlighting the promise of foundation models with self-supervised pretraining in clinical imaging tasks.
\end{abstract}


\section{Introduction}
Retinal vasculitis (RV) is an ocular inflammatory condition characterized by retinal vessel leakage and/or occlusion. RV is mostly idiopathic and $80-90\%$ of cases are non-infectious. It has high association with systemic autoimmune diseases such as Behçet’s disease, systemic lupus erythematosus (SLE), and granulomatosis with polyangiitis. The presence of RV, via vascular leakage, is a sight-threatening condition which leads to refractory cases and worsens the visual prognosis in patients with ocular inflammation. Fluorescein angiography (FA)~\cite{FA_method}, especially with a wide-angle technology, is currently the gold-standard imaging in detecting RV, providing high-resolution visualization of vascular leakage even at subclinical level. The detection of vascular leakage on FA is a hallmark sign of active ocular inflammation and serves as a crucial biomarker for disease monitoring and treatment response. Figure 1 shows representative FA images of the non-infectious RV and its corresponding ground truth masks. These examples illustrate the challenges involved in segmenting retinal vascular leakage regions. Most leakage areas are small in size and densely distributed, requiring the expertise of experienced clinicians to ensure accurate diagnosis, severity assessment and monitoring. Though several FA scoring methods have been described in the literature to standardize assessment, all of them are clinician dependent and require manual annotation~\cite{karaca2023, karaca2024, karaca2025}.

\begin{figure}[t]
    \centering
    \includegraphics[width=0.8\linewidth]{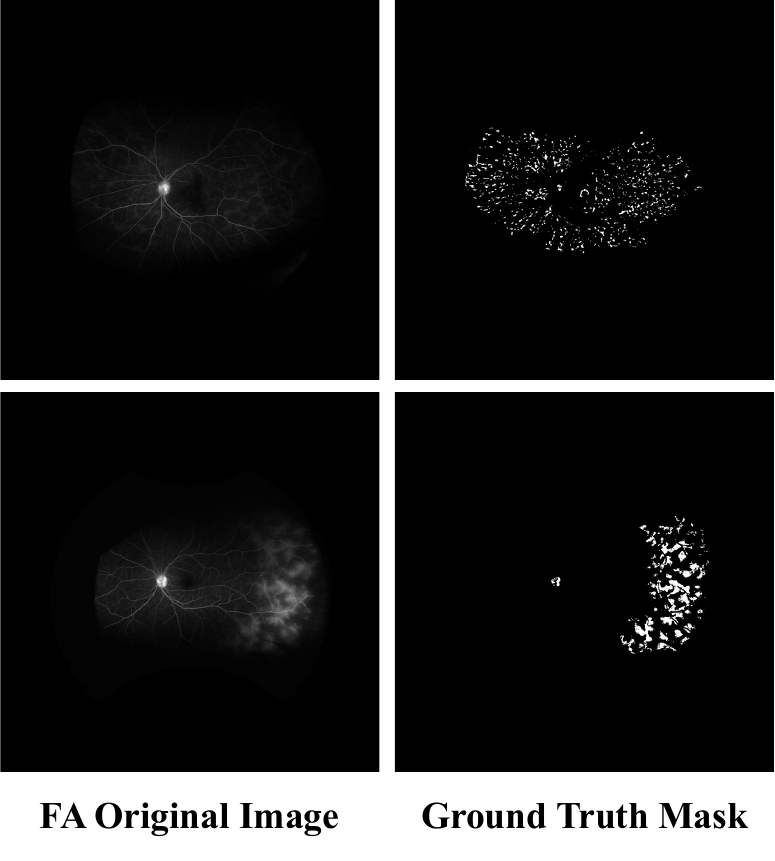}
    \caption{The illustration of the FA original images and ground truth masks. Most leakage areas are small in size and densely distributed, a significant challenge for manual annotation.}    
    \label{fig:instance}
\end{figure}

Currently, AI for healthcare, especially deep learning model, has shown the tremendous potential in medical imaging analysis including the medical imaging segmentation. There are several deep learning applications on different medical image segmentation, such as brain~\cite{brats}, retina vessel~\cite{drive}, and abdomen~\cite{lits}. Unfortunately, unlike to other segmentation tasks in other organ systems, very limited work related to the segmentation of the vascular leakage in RV. To our best knowledge, we only find one work~\cite{RV}  related to our task. The reason is that no labeled dataset is available, and the mask annotation is very difficult and expensive. Considering the sight-threatening nature, requirement of FA imaging as a gold-standard assessment and delicate changes on FA that could be best assessed by subspeciality trained ophthalmologists, standardized  evaluation of RV is highly important. Moreover, given possible FA imaging scan artifacts (e.g. overexposure), the segmentation of the vascular leakage in RV is a challenge.  Our initial attempts, following the work~\cite{RV}, based on the classical medical segmentation models (e.g. Unet~\cite{unet} and DeepLabV3+~\cite{deeplabv3+}), cannot achieve the satisfied performance. 

Another major challenge of the index study lies in the limited dataset size. Training deep learning models for medical imaging tasks typically requires a substantial amount of annotated data for effective supervision. As previously mentioned, clinically labeled datasets are often scarce due to the high cost and time required for expert annotations. Although data augmentation techniques~\cite{augmentation} can help increase data variability and partially mitigate overfitting, they cannot fully overcome the limitations posed by insufficient training data.

To solve the aforementioned challenges, we propose our model MAE-SAM2, a Self-Supervised Learning (SSL) segmentation foundation model. Foundation model is a large-scale and large dataset pre-trained model, has strong generalization capacity over different tasks, can be adapted to different downstream tasks with minimal task-specific modifications. Recent work on foundation models, such as GPT3~\cite{GPT3} of natural language of processing, DINO model~\cite{DINO} of computer vision, show outstanding performance. Segment Anything Model 2 (SAM2)~\cite{SAM2} is a segmentation foundation model proposed by META AI in 2024, a upgraded version of SAM~\cite{SAM}. SSL is a representation learning framework that apply the attributes of the data itself as the supervision during the pretext training, without requiring the manually annotation. Through the SSL pretext task learning stage, the deep learning model is about to learn and understand the semantic and content of the data. Many works have proved the effectiveness and improvement of SSL application in medical data analysis~\cite{context, rotation, permutation}, especially for the limited data.  

Except for the deep learning model design, loss function plays a critical role for the model weights optimization and directly impacts the model segmentation performance. In our task, the vascular leakage regions are typically small, delicate, densely distributed, and randomly located across the image. Moreover, there is a significant class imbalance between foreground (leakage) and background regions, which poses additional challenges for accurate segmentation. To address these issues, we explored several loss function designs and proposed a combined loss strategy to better optimize model training. This approach resulted in more robust and satisfactory segmentation performance across challenging clinical cases.

In summary, our contributions are three-fold:
\begin{itemize}
\item We propose MAE-SAM2, a novel SSL-based segmentation foundation model for retina vascular segmentation on FA image. To the best of our knowledge, this is the first work to apply a foundation model to the retinal vascular leakage segmentation task.
\item We design a combined loss in our task. By investigating various loss function formulations, we proposed a combined loss strategy to enhance model optimization and improve segmentation performance. 
\item We demonstrate the superior performance of MAE-SAM2 model. We conducted a series of experiments and compared our approach against several baseline models. The results demonstrate that our model achieves state-of-the-art (SoTA) performance in retinal vascular leakage segmentation.  

\end{itemize}
\section{Related Works}

\subsection{Deep Learning Models}
Several deep learning models have been developed for medical image segmentation. Among them, U-Net~\cite{unet} is one of the most well-known architectures. It features a U-shaped convolutional neural network (CNN) design with skip connections, which effectively preserve spatial information across the network. Over time, U-Net has evolved into a family of architectures, giving rise to variants such as U-Net++~\cite{unet++} and nnU-Net~\cite{nnunet}, many of which are widely adopted in medical segmentation tasks or used as baselines for benchmarking new models. Another prominent segmentation framework is the encoder-decoder architecture, for example DeepLabV3+~\cite{deeplabv3+}. This model captures multi-scale contextual information through its encoder and reconstructs the segmentation mask via its decoder. In terms of model architecture, self-attention mechanism has gained popularity, particularly with the emergence of Vision Transformers (ViT). These models replace convolutional layers with self-attention blocks to capture global dependencies, leading to models such as SegFormer\cite{segformer} and SwinUnet\cite{swin-unet}, which have demonstrated strong performance in general computer vision tasks. However, in our task, we observed that these existing deep learning models are unsuitable, especially in handling small, densely distributed leakage regions in retinal images. This limitation motivated us to design a novel architecture better suited for the unique challenges of our application.  

\subsection{Foundation Models}
The remarkable success of foundation models in natural language processing and computer vision have sparked growing interest in their application to medical imaging~\cite{gu2024build, sam-med2d}. One prominent research direction involves training foundation models across diverse medical modalities such as MRI and X-ray, thereby enhancing their generalization capabilities across various clinical domains~\cite{gu2024build, sam-med2d}. Another focus has been on developing efficient adaptation methods for these models, including prompt tuning and lightweight fine‑tuning strategies~\cite{cemb}. Recently, researchers have explored zero‑shot inference using foundation models on medical data, evaluating robustness and using prompt augmentation techniques to reduce supervision requirements~\cite{AxonCallosumEM}. Our work similarly builds on the foundation model paradigm: we employ SAM2, a segmentation foundation model, adapted for a specific clinical segmentation task.

\subsection{Self-Supervised Learning}
Many studies have adopted self-supervised learning (SSL) strategies in medical imaging scenarios to improve model performance~\cite{context, rotation, permutation}. Based on the pretext task categories, SSL approaches have the predictive task and generative task. The predictive task involves learning the intrinsic attributes of the data itself, such as context prediction~\cite{context}, rotation prediction~\cite{rotation}, and patch permutation prediction~\cite{permutation}. For the generative tasks, studies focus on reconstructing or denoising the input, including denoising autoencoder~\cite{denoise} and masked image reconstruction~\cite{mae}. Given the dataset may have multi-modality resource, the contrastive learning methods~\cite{SimCLR, MoCo} is used to learn the generalizable representations by maximizing similarity between different modalities of the patient while distinguishing from others. Our work adopts the Masked Autoencoder (MAE), a generative method to improve the SAM2 performance.  

\subsection{Loss function}
Loss function plays a critical role in optimizing the deep learning model for medical data segmentation. Standard loss function like Binary Cross Entropy (BCE)~\cite{bce} and Dice Loss~\cite{dice} have been widely adopted due to their effectiveness in handling pixel-wise classification and overlap-based evaluation, respectively. Medical images often face the small foreground within large background areas,  causing the data imbalance problem. Several works have proposed different loss functions to handle this issue, such as Focal Loss~\cite{focal}, Tversky Loss~\cite{tversky} or Focal Tversky Loss~\cite{focal_tversky}. Additionally, Weighted BCE Loss~\cite{weighted_bce} introduces pixel-level weighting strategies to improve model sensitivity in detecting small regions. Recent studies~\cite{focal} also explore combined loss by integrating multiple functions to guide model optimization more effectively. In our study, we evaluate several of these popular loss functions and design a Combined Loss strategy to enhance segmentation performance on retinal vascular leakage.

\section{Approaches}
We first provide a detailed description of Segmentation Anything Model 2 (SAM2) followed by a brief overview of the Masked Autoencoder (MAE). We then present the overall model architecture and conclude with a discussion of the combined loss function.

\subsection{SAM2 Architecture}

\begin{figure}[htbp]
    \centering
    \includegraphics[width=\linewidth]{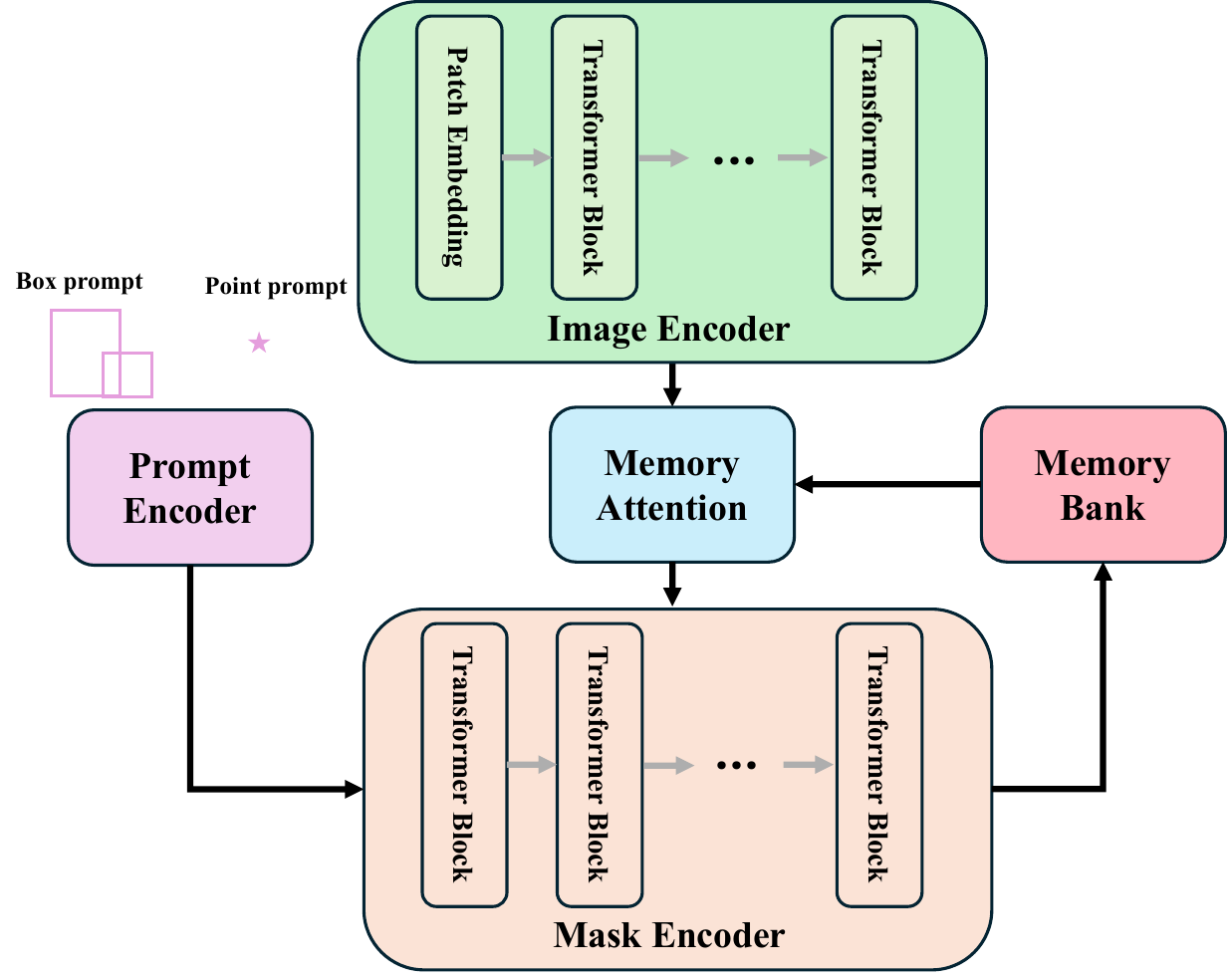}
    \caption{Overview of the SAM2 architecture. SAM2 consists of four main components: an image encoder for extracting visual features, a prompt encoder for interpreting user-provided prompts, a memory module for storing and updating frame information, and a mask decoder for generating segmentation masks.}
    \label{fig:sam2_architecture}
\end{figure}

Proposed by Meta AI in 2024, the SAM2 model is a  promptable visual segmentation foundation model builds upon and extends the capabilities of the original SAM model. Using a powerful data engine, a novel architectural design, and large-scale training on 50.9K videos and 35.5 million corresponding masks, SAM2 demonstrates strong generalization capabilities across video and image segmentation tasks. As illustrated in Figure~\ref{fig:sam2_architecture}, the SAM2 architecture comprises four main components: an image encoder, a memory module, a prompt encoder, and a mask decoder.

In terms of the model implementation, both the image encoder and mask decoder adopt vision transformer (ViT) structures, built upon the transformer blocks. Considering the depth number of the image encoder, the SAM2 has four structure variants: Tiny, Small, Base, and Large model sizes. SAM2 utilizes a prompt encoder to interpret user inputs and guide the segmentation process. The promptable inputs include box prompt, point prompt, mask prompt, and none prompt. The first three types enable interactive segmentation by allowing users to specify regions of interest, while the none prompt corresponds to a fully automatic segmentation mode. Compared with the SAM model architecture, a key innovation of SAM2 is the integration of a memory module, which stores key representations from previously processed frames, allowing the model to efficiently reference and align past contextual information when segmenting the current frame for video segmentation. By maintaining a dynamic memory bank of image features and associated prompts, the memory module improves the model’s capacity to track and segment objects over long-term video understanding tasks. Once the input is the image, the function capacity of the SAM2 model equals to the SAM model.  However, due to the updated architecture design, SAM2 model has four times efficiency over SAM model on the same image segmentation task. Therefore, in our model foundation model backbone choice, we adopt the SAM2 model.

\subsection{MAE}

\begin{figure}[t]
    \centering   \includegraphics[width=0.9\linewidth]{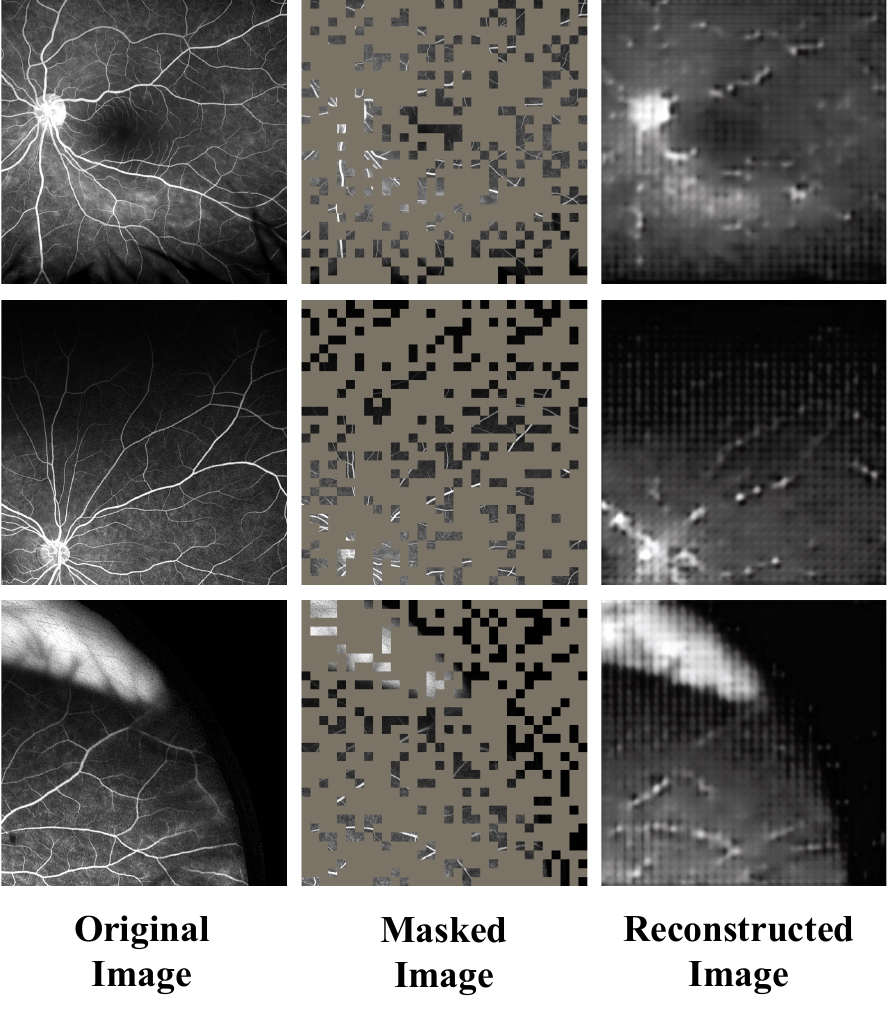}
    \caption{Visual examples of the MAE reconstruction process. From left to right: the original input image, the masked image with a $75\%$ mask ratio, and the corresponding reconstructed image.}
    \label{fig:mae_instance}
\end{figure}

\begin{figure*}[t]
    \centering
    \includegraphics[width=\linewidth]{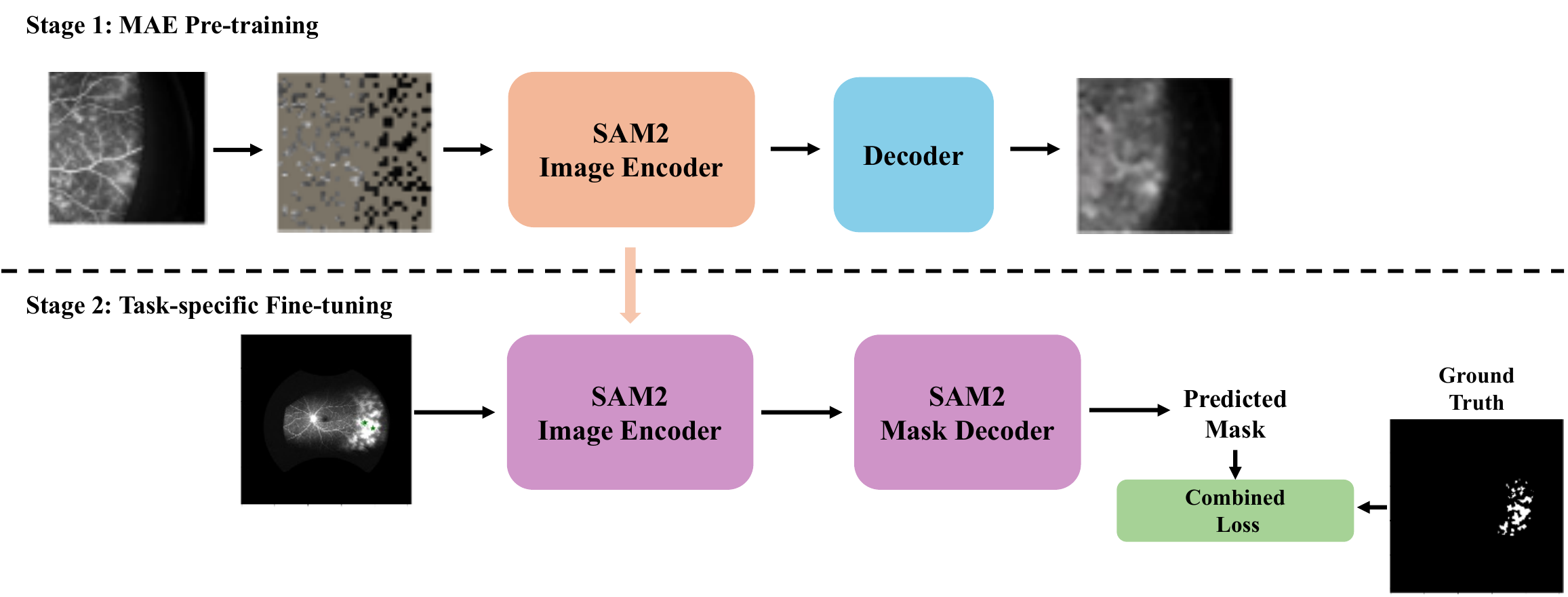}
    \caption{The overview of the MAE-SAM2 architecture. The whole framework has two stages: pre-training stage and fine-tuning stage.}
    \label{fig:architecture}
\end{figure*}

Masked Autoencoder (MAE)\cite{mae} is a self-supervised learning (SSL) framework designed to help deep learning models better understand downstream tasks and enhance performance. Its core idea is to treat image reconstruction as the pretext SSL task, using the image itself as supervision during training. The training process is straightforward: a high proportion of input image patches are randomly masked, and the remaining unmasked patches are fed into a Vision Transformer-based encoder. A lightweight decoder then is trained to reconstruct the original image from this partial input tokens. The mask ratio, a key hyperparameter in MAE, controls the level of reconstruction difficulty.

In the original work, the authors propose a high mask ratio of 0.75, demonstrating that such a challenging setting yields effective self-supervised representations. Given the domain differences between natural and medical images, we evaluated multiple mask ratios (0.25, 0.5, and 0.75) to assess their impact on model performance. Figure\ref{fig:mae_instance} illustrates representative instances showing the original input, the masked images at 0.75 mask ratio, and their corresponding reconstruction outputs. Based on the visual results, although the reconstructed images appear blurred, a visual comparison with the original inputs demonstrates that MAE pretraining contributes the model’s capacity to capture structural and semantic information in medical imaging data.

\subsection{Model Architecture}
\label{model_arch}

Figure~\ref{fig:architecture} illustrates the the overall framework of MAE-SAM2 model. As described earlier, Meta AI offers four SAM2 model architectures: Tiny, Small, Base, and Large. Given the limited size of our dataset, we adopt the Tiny and Small variants of SAM2 as the backbone in our experiments. The MAE-SAM2 training process follows a self-supervised learning (SSL) paradigm and consists of two stages: MAE pretraining and Task-specific fine-tuning.

\textbf{MAE pre-training stage:} In this stage, we concentrate pretraining on the SAM2 image encoder. Following the MAE processing, original input images first are processed through the randomly masked operation. The masked images are then forwarded through the SAM2 image encoder to extract features  and generate a latent feature embedding. The lightweight decoder reconstructs the whole image based on the feature embedding with the original as the supervision. In our design, the decoder consists of four identical blocks, each containing a convolution layer with kernel $3 \times 3$, followed by a batch normalization layer with RELU activation function, and an upsampling layer with scaling factor of 2. The pre-training procedure guides the SAM2 image encoder to learn meaningful representations of FA images by updating its weights. These pretrained weights are subsequently used to initialize the encoder during the fine-tuning stage.

\textbf{Task-specific Fine-tuning stage:} In this stage, we optimize the whole SAM2 model for our specific clinical retina vascular leakage segmentation task. During the training, the input FA image and its corresponding ground truth mask are used as supervision. The image is passed through the MAE-pretrained SAM2 image encoder and the SAM2 mask decoder to generate a predicted segmentation mask. A combined loss function is then computed between the predicted mask and the ground truth annotation, enabling end-to-end fine-tuning of the entire SAM2 model to improve the performance on this domain-specific task.

\subsection{Combined Loss}
Due to the sparse and irregular nature of the vascular leakage masks, clinical retinal vascular segmentation is a challenging task. The vascular leakage masks are small lesions with uncertain or fuzzy boundaries. Furthermore, the foreground and background pixel distribution in FA image is highly imbalanced, causing high false positive and false negative rates in segmentation prediction. To address these challenges, we adopt a combined loss $\mathcal{L}_{\text{Combined}}$ that integrates Dice Loss $\mathcal{L}_{\text{Dice}}$ and Binary Cross-Entropy (BCE) $\mathcal{L}_{\text{BCE}}$. This combined loss leverages both losses to guide the model toward better small lesion segmentation and overall segmentation accuracy.

Dice loss is widely used in segmentation tasks to optimize the overlap between predicted and ground truth masks. It is particularly effective for evaluating the overall shape and structure of the segmented region. Let's present the predicted mask as notation $P = \{ p_{i}\}$ and the ground truth mask as notation $G= \{ g_{i}\}$. The Dice Loss $L_{Dice}$ definition is following:

\begin{equation}
\mathcal{L}_{\text{Dice}} = 1 - \frac{2 |P \cap G|}{|P| + |G| }
\end{equation}

 Dice loss can be less sensitive to small objects or hard-to-segment regions, especially when the class distribution is highly imbalanced.  To address these limitations, we incorporate Binary Cross-Entropy (BCE) Loss, which provides fine-grained supervision at the pixel level and complements Dice Loss in handling challenging segmentation cases:

\begin{equation}
\mathcal{L}_{\text{BCE}} = - \left[ G \cdot \log(P) + (1 - G) \cdot \log(1 - P) \right]
\end{equation}

The final loss function is a weighted sum of the two components:
\begin{equation}
\mathcal{L}_{\text{combined}} = \lambda_1 \cdot \mathcal{L}_{\text{Dice}} + \lambda_2 \cdot \mathcal{L}_{\text{BCE}}
\end{equation}

where $\lambda_1$ and $\lambda_2$ are tunable hyperparamterin our implementation.

\section{Results}
\subsection{Dataset and Metrics}
The dataset used in this study consists of privately collected clinical fundus fluorescein angiography (FA) scans. Ground truth masks were manually annotated by clinical experts to ensure high-quality supervision. In total, the dataset comprises 74 FA images obtained from 38 patients, with each image having a resolution of $3096 \times 3096$ pixels. We performed an 80\% vs. 20\% subject-level split, assigning 28 patients (53 scans) to the training set and 10 patients (21 scans) to the test set. Given the limited dataset size, we applied a patch-based augmentation strategy to enlarge the training set. Specifically, we extracted image patches of size $1024 \times 1024$ with a stride of 512 pixels from the training images. Patch extraction was performed exclusively on the training set. For the test set, the original images were resized to $1024 \times 1024$ for inference and evaluation. The training configuration uses a batch size of 16 for 50 epochs, with the AdamW optimizer and an initial learning rate of $1.0 \times 10^{-4}$, scheduled via the \texttt{LambdaLR} scheduler.

To evaluate the model performance on our task, we choose the following metrics: Dice Score, Intersection-over-Union (IOU), Precision, Recall, F1 Score, Specificity. \textbf{We adopt the Dice Score as the primary metric for model performance evaluation}. 

\subsection{Model Performance}

\begin{figure}[htbp]
    \centering
    \includegraphics[width=0.9\linewidth]{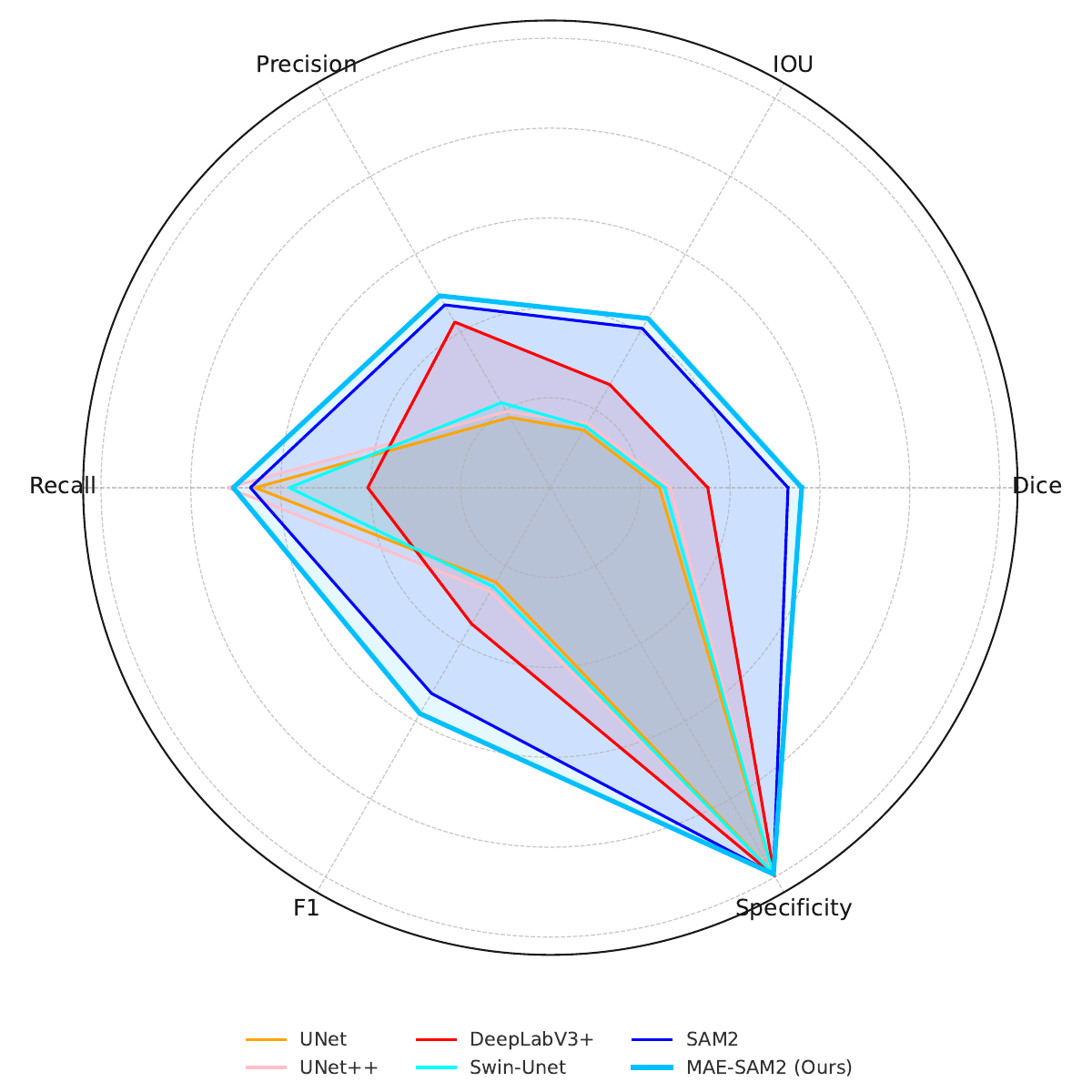}
    \caption{The radar graph visualization of different model performance over the different models. The MAE-SAM2 outperforms the baselines over different metrics.}
    \label{fig:model_performance}
\end{figure}

\begin{table*}[ht]
\caption{Comparison of Segmentation Performance Across Different Models}
\begin{center}
\renewcommand{\arraystretch}{1.3}  
\begin{tabular}{|>{\centering\arraybackslash}p{2.4cm}|
    >{\centering\arraybackslash}p{1.8cm}|
    >{\centering\arraybackslash}p{1.8cm}|
    >{\centering\arraybackslash}p{1.8cm}|
    >{\centering\arraybackslash}p{1.8cm}|
    >{\centering\arraybackslash}p{1.8cm}|
    >{\centering\arraybackslash}p{1.8cm}|
    }
\hline
\textbf{Model} & \textbf{Dice Score} & \textbf{IOU} & \textbf{Precision} & \textbf{Recall} & \textbf{F1 Score} & \textbf{Specificity}  \\
\hline
UNet~\cite{unet} &0.2426  &0.1481  &0.1803  &0.6567  &0.2426  &0.9861 \\
UNet++~\cite{unet++} &0.2668 & 0.1655  & 0.1955 & 0.7182 & 0.2668 & 0.9855  \\
DeepLabV3+~\cite{deeplabv3+} &0.3502  &0.2647  &0.4256  &0.4063  &0.3502  &0.9972  \\
Swin-Unet~\cite{swin-unet} &  0.2543 & 0.1573  & 0.2183 & 0.5809 & 0.2543 & 0.9908 \\
MedSAM2~\cite{SAM2} & 0.5288 & 0.4092  & 0.4696  & 0.6675 & 0.5288   &0.9927 \\
MAE-SAM2 (\textbf{Ours}) &  \textbf{0.5593} & \textbf{0.4348} & \textbf{0.4932} & \textbf{0.7055} & \textbf{0.5805} & \textbf{0.9929} \\
\hline
\end{tabular}
\label{tab:seg_comparison}
\end{center}
\end{table*}

We compare the MAE-SAM2 model performance with several baselines; they are well-known and frequently used deep learning segmentation models: 
\begin{itemize}
    \item Unet~\cite{unet} is a classical and famous baseline of the convolutional neural network (CNN) segmentation model.
    \item Unet++~\cite{unet++}  is a varients of the Unet by redesigning the skip connections using nested and dense convolutional blocks.
    \item DeepLabV3+~\cite{deeplabv3+} is a SOTA semantic segmentation model that incorporates dilated convolution layers and an Atrous Spatial Pyramid Pooling (ASPP) module to capture multi-scale contextual information.
    \item SwinUnet~\cite{swin-unet} is a transformer-based segmentation architecture that adapts the Swin Transformer as its backbone.
\end{itemize}

Figure~\ref{fig:model_performance} illustrates the segmentation performance across different models in a radar plot format, while Table~\ref{tab:seg_comparison} presents the detailed quantitative comparison of their performance metrics. Based on the quantitative experiments, the CNN-based architectures such as U-Net~\cite{unet}, U-Net++~\cite{unet++}, and DeepLabV3+~\cite{deeplabv3+} demonstrate limited effectiveness on our clinical dataset, with Dice Scores of 0.2426, 0.2668, and 0.3502, respectively. Swin-Unet~\cite{swin-unet}, a transformer-based model, also shows unsatisfied performance with a Dice Score of 0.2543. The original SAM2 model achieves significantly better results, with a Dice Score of 0.5288, proving the benefit of segmentation foundation model. Our proposed MAE-SAM2 model further improves performance, achieving the highest Dice Score (0.5595), IoU (0.4350), Precision (0.5293), and Specificity (0.9952) among all compared models. Compared with original SAM2 model, MAE-SAM2 enhances a $5.8\%$ improvement in Dice Score Metric. These results highlight the effectiveness of our proposed MAE-SAM2 model, by adopting the self-supervised MAE pretraining in enhancing the SAM2 encoder's representation learning, especially under the constraints of limited data and challenging clinical features.

\subsection{Experiments on Hyperparameters of SAM2}

\begin{table}[h]
\caption{Evaluation of SAM2 hyperparameters, including model architecture variants (Tiny and Small) and prompt types (point, box, and none), and their effects on segmentation performance.}
\begin{center}
\renewcommand{\arraystretch}{1.3}
\begin{tabular}{| >{\centering\arraybackslash}p{2.4cm}  >{\centering\arraybackslash}p{2.0cm} | >{\centering\arraybackslash}p{2.4cm} |}
\hline
\textbf{Model} & \textbf{Prompt} & \textbf{Dice Score} \\
\hline
Tiny & point & 0.5222 \\
Tiny & box & 0.4161 \\
Small & none & 0.5023 \\
Small & box & 0.5130 \\
Small & point & \textbf{0.5288} \\
\hline
\end{tabular}
\label{tab:sam2_comparison}
\end{center}
\end{table}

In this section, we investigate the impact of SAM2's model architecture and prompt types on segmentation performance. As previously described, SAM2 offers four architecture variants: Tiny, Small, Base, and Large. Due to limitations in dataset size and computational resources, we focus on the SAM2-Tiny and SAM2-Small architectures in our implementation. Additionally, SAM2 supports various prompt types, including point, box, mask, and none. In our experiments, we compare three prompt inputs: point, box, and none, to evaluate their effectiveness in guiding the segmentation process. 

Table~\ref{tab:sam2_comparison} reports the Dice scores under different SAM2 configurations. For the same SAM2-Small architecture, the point prompt achieves superior performance compared to the box and none prompts, which meet our expectations.  This is likely because point prompts effectively provide localized regions of interest to guide the model toward learning the segmentation task more precisely. The results indicate that the SAM2-Small architecture achieves better performance than SAM2-Tiny, suggesting that large model capacity contributes to improved segmentation. Therefore, in out final model design, we adopt SAM2-Small as the backbone and apply point prompt during training and evaluation.

\begin{figure*}[t]
    \centering
    \includegraphics[width=\linewidth]{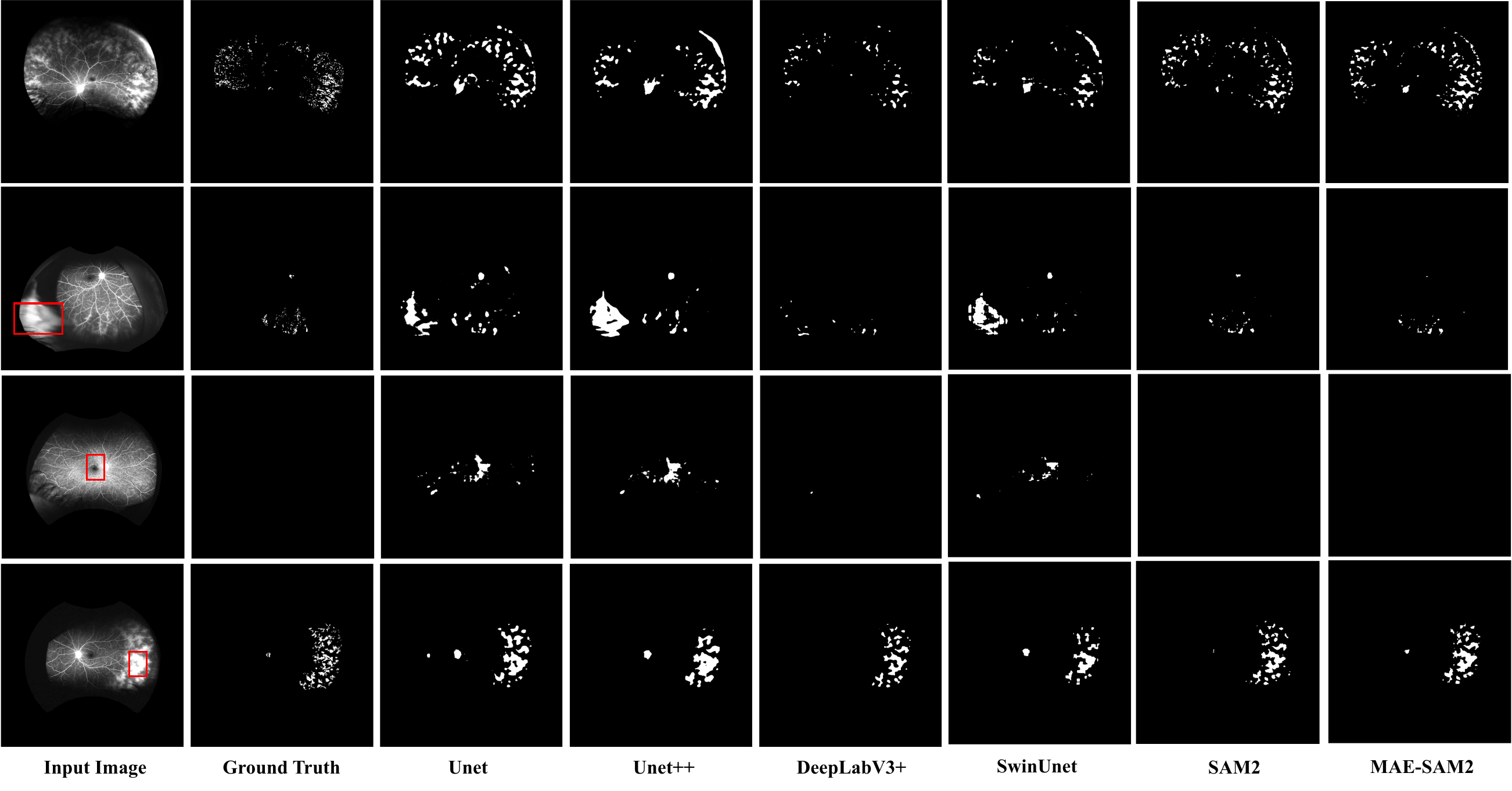}
    \caption{Visualization of segmentation results produced by different models. Each row shows a sample input image, its corresponding ground truth mask, and the predicted segmentation masks from UNet, UNet++, DeepLabV3+, SwinUnet, SAM2, and the proposed MAE-SAM2.}
    \label{fig:visulization}
\end{figure*}

\subsection{Loss Function Comparison}

\begin{table}[htbp]
\caption{Evaluation on different loss functions and their effects on MAE-SAM2 model performance.}
\begin{center}
\renewcommand{\arraystretch}{1.3}
\begin{tabular}{| >{\centering\arraybackslash}p{3.0cm} | >{\centering\arraybackslash}p{2.0cm}  >{\centering\arraybackslash}p{2.4cm} |}

\hline
\textbf{Loss Type} & \textbf{Dice Score} & \textbf{IOU} \\
\hline
Dice Loss~\cite{dice} & 0.5438 & 0.4228 \\
Focal Loss~\cite{focal} & 0.5368 & 0.4162 \\
Focal Tversky Loss~\cite{focal_tversky} & 0.5260 & 0.4056 \\
BCE Loss~\cite{bce} & 0.5376 & 0.4172 \\
Weighted BCE Loss~\cite{weighted_bce} & 0.5272 & 0.4073 \\
Combined Loss & \textbf{0.5593} & \textbf{0.4348}\\

\hline
\end{tabular}
\label{tab:loss_comparison}
\end{center}
\end{table}

In this section, we explore different loss functions in our task. We conducted a series of experiments to guide the selection of the final loss function. Several loss functions have been widely used in medical image segmentation. In our experiments, we evaluated the following loss functions: Dice Loss~\cite{dice}, Focal Loss~\cite{focal}, Focal Tversky Loss~\cite{focal_tversky}, Binary Cross-Entropy (BCE) Loss~\cite{bce}, and Weighted BCE Loss~\cite{weighted_bce}. Each loss function was independently evaluated under the same experimental settings to ensure a fair comparison. Based on the performance of individual losses, we further explored a combined loss strategy to determine the optimal formulation. We began with Dice Loss as the baseline and evaluated its combinations with BCE Loss and Focal Tversky Loss. Experimental results indicate that the most effective configuration is a combination of Dice Loss and BCE Loss. Further coefficient tuning reveals that assigning a weight of 0.9 to Dice Loss and 0.1 to BCE Loss yields the best segmentation performance. Table~\ref{tab:loss_comparison} presents the performance of the model under different loss functions, where the proposed combined loss achieves the best overall results.

\subsection{Ablation Study}

\begin{table}[htbp]
\caption{Evaluation on different loss functions and their effects on MAE-SAM2 model performance.}
\begin{center}
\renewcommand{\arraystretch}{1.3}
\begin{tabular}{| >{\centering\arraybackslash}p{2.4cm}  >
{\centering\arraybackslash}p{2.0cm} | >{\centering\arraybackslash}p{2.4cm} |}

\hline
\textbf{MAE} & \textbf{Combined Loss} & \textbf{Dice Score} \\
\hline
            &             & 0.5288 \\
            & \checkmark  & 0.5305 \\
\checkmark  &             & 0.5438 \\
\checkmark  & \checkmark  & 0.5593 \\
\hline
\end{tabular}
\label{tab:ablation}
\end{center}
\end{table}

We conducted a simple ablation study to evaluate the contribution of each component in the MAE-SAM2 model. As shown in Table~\ref{tab:ablation}, both MAE pretraining and the proposed combined loss play important roles in improving segmentation performance. The baseline SAM2 model, without MAE pretraining or combined loss, achieves a Dice score of 0.5288. Applying the combined loss alone yields a slight improvement, while enabling only MAE pretraining increases the Dice score to 0.5438. When both components are used together, the model achieves the highest Dice score of 0.5593, demonstrating that MAE pretraining and the combined loss jointly enhance the overall performance of the MAE-SAM2 model.

\subsection{Visualization}

Figure~\ref{fig:visulization} illustrates the segmentation results predicted by different models. From left to right, we present the input images, corresponding ground truth masks, and predictions from UNet, UNet++, DeepLabV3+, SwinUNet, SAM2, and MAE-SAM2. Several observations can be made from these visual examples. Overall, SAM2 and MAE-SAM2 demonstrate superior segmentation performance compared to other baseline models, highlighting the potential of promptable segmentation foundation models in clinical medical imaging tasks. In particular, vascular leakage segmentation is a small-lesion task, and both SAM2 and MAE-SAM2 exhibit stronger capabilities in capturing these fine-grained pathological features.

During FA image scanning, overexposure artifacts are commonly observed (highlighted by red bounding boxes in Figure~\ref{fig:visulization}), which degrade image quality and introduce challenges for accurate segmentation. Most baseline models fail to address these artifacts—misclassifying overexposed regions as leakage areas (e.g., in rows 2, 3, and 4). Notably, the image in row 3 is from a healthy subject, yet several baseline models produce numerous false positive predictions. In contrast, both SAM2 and MAE-SAM2 exhibit robustness to such noise and produce more accurate results. Although the visual difference between SAM2 and MAE-SAM2 may not be obvious in every case, quantitative metrics such as Dice Score and IoU consistently show that MAE-SAM2 outperforms SAM2 across the entire test set.

\section{Conclusion}
In this work, we proposed MAE-SAM2 for retinal vascular leakage segmentation, the first foundation model on this specific clinical task. Our design addresses key challenges in clinical retinal vascular leakage segmentation, including limited data and small lesion size. Experimental results show that MAE-SAM2 achieves superior performance over existing baselines, proving the potential of foundation model and SSL on clinical data. Our future work will include the foundation model efficient tuning and expanding the current work with prompt learning.


\begin{thebibliography}{00}

\bibitem{FA_method}
Harold R.~Novotny and David L.~Alvis.
\newblock A method of photographing fluorescence in circulating blood in the human retina.
\newblock \emph{Circulation}, 24(1):82--86, 1961. Lippincott Williams \& Wilkins.

\bibitem{karaca2025}
Irmak Karaca, et~al.
\newblock Efficacy and safety of biologics in pediatric non-infectious retinal vasculitis.
\newblock \emph{American Journal of Ophthalmology}, 2025. Elsevier.

\bibitem{karaca2024}
Irmak Karaca, et~al.
\newblock Importance of baseline fluorescein angiography for patients presenting to tertiary uveitis clinic.
\newblock \emph{American Journal of Ophthalmology}, 265:296--302, 2024. Elsevier.

\bibitem{karaca2023}
Irmak Karaca, et~al.
\newblock Six-month outcomes of infliximab and tocilizumab therapy in non-infectious retinal vasculitis.
\newblock \emph{Eye}, 37(11):2197--2203, 2023. Nature Publishing Group UK London.


\bibitem{brats}
Bjoern H.~Menze, Andras Jakab, Stefan Bauer, Jayashree Kalpathy-Cramer, Keyvan Farahani, 
Justin Kirby, et~al.
\newblock The Multimodal Brain Tumor Image Segmentation Benchmark (BRATS).
\newblock \emph{IEEE Transactions on Medical Imaging}, 34(10):1993--2024, 2015.

\bibitem{drive}
Josien M.~Staal, Michael D.~Abràmoff, Meindert Niemeijer, Max A.~Viergever, and Bram van Ginneken.
\newblock Ridge-based vessel segmentation in color images of the retina.
\newblock \emph{IEEE Transactions on Medical Imaging}, 23(4):501--509, 2004. (DRIVE dataset)

\bibitem{lits}
Patrick Bilic, Patrick F.~Christ, Eugene Vorontsov, Gabriel Chlebus, Holger Chen, 
Qi Dou, et~al.
\newblock The Liver Tumor Segmentation Benchmark (LiTS).
\newblock \emph{arXiv preprint arXiv:1901.04056}, 2019.


\bibitem{RV}
Dhanach Dhirachaikulpanich, Savita Madhusudhan, David Parry, 
Salma Babiker, Yalin Zheng, and Nicholas A.~V. Beare.
\newblock Retinal vasculitis severity assessment: intra- and inter-observer
reliability of a new scheme for grading wide-field fluorescein angiograms
in retinal vasculitis.
\newblock \emph{Retina}, pages 10--1097, 2022. LWW.

\bibitem{unet}
Olaf Ronneberger, Philipp Fischer, and Thomas Brox.
\newblock U-net: Convolutional networks for biomedical image segmentation.
\newblock In \emph{International Conference on Medical Image Computing and Computer-Assisted Intervention}, pages 234--241. Springer, 2015.


\bibitem{deeplabv3+}
Liang-Chieh Chen, Yukun Zhu, George Papandreou, Florian Schroff, and Hartwig Adam.
\newblock Encoder-decoder with atrous separable convolution for semantic image segmentation.
\newblock In \emph{Proceedings of the European Conference on Computer Vision (ECCV)}, pages 801--818, 2018.

\bibitem{unet++}
Zongwei Zhou, Md Mahfuzur Rahman Siddiquee, Nima Tajbakhsh, and Jianming Liang.
\newblock Unet++: A nested u-net architecture for medical image segmentation.
\newblock In \emph{International Workshop on Deep Learning in Medical Image Analysis}, pages 3--11. Springer, 2018.

\bibitem{nnunet}
Fabian Isensee, Jens Petersen, Andre Klein, David Zimmerer, Paul F.~Jaeger, 
Simon Kohl, Jakob Wasserthal, Gregor Koehler, Tobias Norajitra, Sebastian Wirkert, et~al.
\newblock nnu-net: Self-adapting framework for u-net-based medical image segmentation.
\newblock \emph{arXiv preprint arXiv:1809.10486}, 2018.

\bibitem{swin-unet}
Hu Cao, Yueyue Wang, Joy Chen, Dongsheng Jiang, Xiaopeng Zhang, Qi Tian, and Manning Wang.
\newblock Swin-unet: Unet-like pure transformer for medical image segmentation.
\newblock In \emph{European Conference on Computer Vision}, pages 205--218. Springer, 2022.

\bibitem{augmentation}
Connor Shorten and Taghi M.~Khoshgoftaar.
\newblock A survey on image data augmentation for deep learning.
\newblock \emph{Journal of Big Data}, 6(1):60, 2019.

\bibitem{GPT3}
Tom B.~Brown, Benjamin Mann, Nick Ryder, Melanie Subbiah, Jared Kaplan, Prafulla Dhariwal, 
Arvind Neelakantan, Pranav Shyam, Girish Sastry, Amanda Askell, Sandhini Agarwal, Ariel Herbert-Voss, 
Gretchen Krueger, Tom Henighan, Rewon Child, Aditya Ramesh, Daniel M.~Ziegler, Jeffrey Wu, 
Clemens Winter, Christopher Hesse, Mark Chen, Eric Sigler, Mateusz Litwin, Scott Gray, 
Benjamin Chess, Jack Clark, Christopher Berner, Sam McCandlish, Alec Radford, Ilya Sutskever, and Dario Amodei.
\newblock Language models are few-shot learners.
\newblock In \emph{Proceedings of the 34th International Conference on Neural Information Processing Systems (NeurIPS)}, pages 1877--1901, 2020.

\bibitem{DINO}
Mathilde Caron, Hugo Touvron, Ishan Misra, Herv{\'e} J{\'e}gou, Julien Mairal, Piotr Bojanowski, and Armand Joulin.
\newblock Emerging properties in self-supervised vision transformers.
\newblock In \emph{Proceedings of the IEEE/CVF International Conference on Computer Vision (ICCV)}, pages 9650--9660, 2021.

\bibitem{SAM}
Alexander Kirillov, Eric Mintun, Nikhila Ravi, Hanzi Mao, Chloe Rolland, Laura Gustafson, 
Tete Xiao, Spencer Whitehead, Alexander C.~Berg, Wan-Yen Lo, et~al.
\newblock Segment anything.
\newblock In \emph{Proceedings of the IEEE/CVF International Conference on Computer Vision}, pages 4015--4026, 2023.

\bibitem{SAM2}
Nikhila Ravi, Valentin Gabeur, Yuan-Ting Hu, Ronghang Hu, Chaitanya Ryali, Tengyu Ma, 
Haitham Khedr, Roman R{\"a}dle, Chloe Rolland, Laura Gustafson, et~al.
\newblock Sam 2: Segment anything in images and videos.
\newblock \emph{arXiv preprint arXiv:2408.00714}, 2024.


\bibitem{segformer}
Enze Xie, Wenhai Wang, Zhiding Yu, Anima Anandkumar, Jose M.~Alvarez, and Ping Luo.
\newblock SegFormer: Simple and efficient design for semantic segmentation with transformers.
\newblock \emph{Advances in Neural Information Processing Systems}, 34:12077--12090, 2021.

\bibitem{gu2024build}
Hanxue Gu, Haoyu Dong, Jichen Yang, and Maciej A.~Mazurowski.
\newblock How to build the best medical image segmentation algorithm using foundation models: a comprehensive empirical study with segment anything model.
\newblock \emph{arXiv preprint arXiv:2404.09957}, 2024.


\bibitem{sam-med2d}
Junlong Cheng, Jin Ye, Zhongying Deng, Jianpin Chen, Tianbin Li, Haoyu Wang, Yanzhou Su,
Ziyan Huang, Jilong Chen, Lei Jiang, Hui Sun, Junjun He, Shaoting Zhang, Min Zhu, and Yu Qiao.
\newblock SAM-Med2D.
\newblock \emph{arXiv preprint arXiv:2308.16184}, 2023.

\bibitem{AxonCallosumEM}
Ao Cheng, Guoqiang Zhao, Lirong Wang, and Ruobing Zhang.
\newblock AxonCallosumEM dataset: Axon semantic segmentation of whole corpus callosum cross section from EM images.
\newblock \emph{arXiv preprint arXiv:2307.02464}, 2023.

\bibitem{cemb}
Dongik Shin, Beomsuk Kim, and Seungjun Baek.
\newblock Cemb-sam: Segment anything model with condition embedding for joint learning from heterogeneous datasets.
\newblock In \emph{International Conference on Medical Image Computing and Computer-Assisted Intervention (MICCAI)}, 
pages 275--284. Springer, 2023.

\bibitem{mae}
Kaiming He, Xinlei Chen, Saining Xie, Yanghao Li, Piotr Doll{\'a}r, and Ross Girshick.
\newblock Masked autoencoders are scalable vision learners.
\newblock In \emph{Proceedings of the IEEE/CVF Conference on Computer Vision and Pattern Recognition}, 
pages 16000--16009, 2022.

\bibitem{context}
Carl Doersch, Abhinav Gupta, and Alexei A.~Efros.
\newblock Unsupervised visual representation learning by context prediction.
\newblock In \emph{Proceedings of the IEEE International Conference on Computer Vision}, 
pages 1422--1430, 2015.

\bibitem{rotation}
Spyros Gidaris, Praveer Singh, and Nikos Komodakis.
\newblock Unsupervised representation learning by predicting image rotations.
\newblock \emph{arXiv preprint arXiv:1803.07728}, 2018.

\bibitem{permutation}
Mehdi Noroozi and Paolo Favaro.
\newblock Unsupervised learning of visual representations by solving jigsaw puzzles.
\newblock In \emph{European Conference on Computer Vision}, pages 69--84. Springer, 2016.

\bibitem{denoise}
Pascal Vincent, Hugo Larochelle, Yoshua Bengio, and Pierre-Antoine Manzagol.
\newblock Extracting and composing robust features with denoising autoencoders.
\newblock In \emph{Proceedings of the 25th International Conference on Machine Learning}, 
pages 1096--1103, 2008.

\bibitem{SimCLR}
Ting Chen, Simon Kornblith, Mohammad Norouzi, and Geoffrey Hinton.
\newblock A simple framework for contrastive learning of visual representations.
\newblock In \emph{International Conference on Machine Learning}, 
pages 1597--1607, 2020. PMLR.

\bibitem{MoCo}
Kaiming He, Haoqi Fan, Yuxin Wu, Saining Xie, and Ross Girshick.
\newblock Momentum contrast for unsupervised visual representation learning.
\newblock In \emph{Proceedings of the IEEE/CVF Conference on Computer Vision and Pattern Recognition}, 
pages 9729--9738, 2020.

\bibitem{bce}
Ian Goodfellow, Yoshua Bengio, and Aaron Courville.
\newblock \emph{Deep Learning}.
\newblock MIT Press, 2016. (Binary Cross-Entropy loss definition, Chapter 6)

\bibitem{dice}
Felix Milletari, Nassir Navab, and Seyed-Ahmad Ahmadi.
\newblock V-Net: Fully convolutional neural networks for volumetric medical image segmentation.
\newblock In \emph{Proceedings of the IEEE International Conference on 3D Vision (3DV)}, pages 565--571, 2016. (Dice Loss)

\bibitem{focal}
Tsung-Yi Lin, Priya Goyal, Ross Girshick, Kaiming He, and Piotr Doll{\'a}r.
\newblock Focal loss for dense object detection.
\newblock In \emph{Proceedings of the IEEE International Conference on Computer Vision (ICCV)}, pages 2980--2988, 2017.

\bibitem{tversky}
Sadegh Salehi, Deniz Erdogmus, and Ali Gholipour.
\newblock Tversky loss function for image segmentation using 3D fully convolutional deep networks.
\newblock In \emph{Proceedings of Machine Learning in Medical Imaging (MLMI)}, pages 379--387. Springer, 2017.

\bibitem{focal_tversky}
Nabila Abraham and Naimul Mefraz Khan.
\newblock A novel focal Tversky loss function with improved attention U-Net for lesion segmentation.
\newblock In \emph{Proceedings of the IEEE International Symposium on Biomedical Imaging (ISBI)}, pages 683--687, 2019.

\bibitem{weighted_bce}
Jingwei Chen, Lequan Yu, Qian Wang, and Pheng-Ann Heng.
\newblock Combining weighted cross entropy loss and Dice loss for medical image segmentation.
\newblock \emph{arXiv preprint arXiv:1802.05140}, 2018.



\end{thebibliography}
\end{document}